\documentclass[fleqn,final,5p,times]{elsarticle}

\usepackage{amssymb}

\usepackage{calc}
\usepackage{siunitx}
\usepackage{physics}
\usepackage[hidelinks]{hyperref}
\usepackage{cleveref}
\usepackage{xcolor}
\usepackage{xargs}
\usepackage{soul}
\usepackage{subcaption}
\usepackage{macros}
\usepackage{booktabs}
\usepackage{tabularx}

\definecolor{lightpurple}{RGB}{  232,211,246}
\definecolor{lightturkis}{RGB}{   71,235,223}

\definecolor{lighterpurple}{RGB}{230,221,236}
\definecolor{lighterturkis}{RGB}{183,255,255}
\definecolor{lightergreen}{RGB}{ 144,255,199}

\definecolor{lightgrey}{RGB}{     80, 90,100}
\definecolor{lightergrey}{RGB}{  130,140,160}
\definecolor{bluegrey}{RGB}{     130,140,180}
\definecolor{whitegrey}{RGB}{    235,240,240}

\definecolor{tabblue}    {HTML}{1f77b4}
\definecolor{taborange}  {HTML}{ff7f0e}
\definecolor{tabgreen}   {HTML}{2ca02c}
\definecolor{tabred}     {HTML}{d62728}
\definecolor{tabpurple}  {HTML}{9467bd}
\definecolor{tabbrown}   {HTML}{8c564b}
\definecolor{tabpink}    {HTML}{e377c2}
\definecolor{tabgray}    {HTML}{7f7f7f}
\definecolor{tabolive}   {HTML}{bcbd22}
\definecolor{tabcyan}    {HTML}{17becf}
\definecolor{tab20blue}  {HTML}{aec7e7}
\definecolor{tab20orange}{HTML}{febb77}
\definecolor{tab20green} {HTML}{97df89}
\definecolor{tab20red}   {HTML}{fe9796}
\definecolor{tab20purple}{HTML}{c5b0d5}
\definecolor{tab20brown} {HTML}{c49b93}
\definecolor{tab20pink}  {HTML}{f6b5d1}
\definecolor{tab20gray}  {HTML}{c6c6c6}
\definecolor{tab20olive} {HTML}{dadb8d}
\definecolor{tab20cyan}  {HTML}{9ed9e5}

\usepackage[mode=buildnew,subpreambles=true]{standalone}
\usepackage{shellesc}
\usepackage{tikz,tikzscale}
\usetikzlibrary{arrows,calc,fit,patterns,positioning,spy,backgrounds,decorations.fractals,shapes.misc}
\usetikzlibrary{shapes.geometric}
\tikzset{boximg/.style={remember picture,black,thick,draw,inner sep=0pt,outer sep=0pt}}
\usepackage{pgf}
\usepackage{pgfplots}
\usepgfplotslibrary{statistics}
\usepgfplotslibrary{fillbetween}
\usepgfplotslibrary{colormaps}
\pgfplotsset{compat=1.18}
\pgfplotsset{lua backend=true}

\pgfplotsset{select coords between index/.style 2 args={
    x filter/.code={
        \ifnum\coordindex<#1\fi
        \ifnum\coordindex>#2\fi
    }
}}

\usepackage{graphicx}
\usepackage{xspace}
\newcommand{\hawk}{\textit{Hawk}\xspace}
\newcommand{\lumi}{\textit{LUMI}\xspace}
\newcommand{\flexi}{FLEXI\xspace}
\newcommand{\elexi}{\reflectbox{E}LEXI\xspace}

\journal{International Journal of \LaTeX templates}

\allowdisplaybreaks

\begin{document}
\sloppy

\begin{frontmatter}

\title{Efficient computation of particle-fluid and particle-particle interactions in compressible flow}

\author[iag]{Anna Schwarz\corref{cor}\fnref{fn}}
\ead{schwarz@iag.uni-stuttgart.de}
\author[iag]{Patrick Kopper\corref{cor}\fnref{fn}}
\ead{kopper@iag.uni-stuttgart.de}
\author[lmfa]{Emilian de Staercke}
\author[iag]{Andrea Beck}

\cortext[cor]{Corresponding author}
\fntext[fn]{P. Kopper and A. Schwarz share first authorship.}

\affiliation[iag]{organization={University of Stuttgart, Institute of Aerodynamics and Gas Dynamics},
            addressline={Pfaffenwaldring 21},
            city={Stuttgart},
            postcode={70569},
            country={Germany}}
\affiliation[lmfa]{organization={École Centrale de Lyon, Laboratoire de Mécanique des Fluides et d'Acoustique},
            addressline={36 Avenue Guy de Collongue},
            city={Écully Cedex},
            postcode={69134},
            country={France}}

\begin{abstract}
Particle collisions are the primary mechanism of inter-particle momentum and energy exchange for dense particle-laden flow.
Accurate approximation of this collision operator in four-way coupled Euler--Lagrange approaches remains challenging due to the associated computational cost.
Adopting a deterministic collision model and a hard-sphere (binary collision) approach eases time step constraints but imposes
non-locality on distributed memory architectures, necessitating the inclusion of collision partners from each grid element in the vicinity.
Retaining high-order accuracy and parallel efficiency also ties into the correct and compact treatment of the particle-fluid coupling, where adequate kernels are required to
effectively project the work of the particles to the Eulerian grid.
In this work, we present an efficient particle collision and projection operator based on an MPI+MPI hybrid approach to enable
time-resolved and high-order accurate simulations of compressible, four-way coupled particle-laden flows at dense concentrations.
A distinct feature of the proposed particle collision algorithm is the efficient calculation of exact binary inter-particle
collisions on arbitrary core counts by facilitating intranode data exchange through direct load/store operations and internode communication using
one-sided communication.
Combining the particle operator with a hybrid discretization operator based on a high-order
discontinuous Galerkin method and a localized low-order finite volume operator allows an accurate treatment of highly compressible particle-laden flows.
The approach is extensively validated against a range of benchmark problems. Contrary to literature, the scaling properties are
demonstrated on state-of-the-art high performance computing systems, encompassing one-way to four-way coupled simulations.
Finally, the proposed algorithm is compatible with unstructured, curved high-order grids which permits the handling of complex
geometries as is emphasized by application of the framework to large-scale application cases.
\end{abstract}

\begin{keyword}
high-order \sep discontinuous Galerkin \sep high-performance computing \sep large eddy simulation \sep Lagrangian particle
tracking \sep particle collisions

\end{keyword}

\end{frontmatter}

\section{Introduction}%
\label{sec:introduction}

Compressible flow with suspended solid particles is present in a wide range of technical applications, ranging from spray injection
to aerospace engineering~\cite{Balachandar2010,Varaksin2023}.
Most of these applications are challenging to simulate due to the strongly non-linear, multi-scale, and multi-physics nature of the problems.
These challenges have led to the development of numerous modeling approaches over the last years, aimed at enabling high-fidelity simulations of such particle-laden flows.
One prominent approach is the Euler--Lagrange or point particle method, in which the particles are described as discrete points.
The particles evolve in a Lagrangian manner according to Newton's second law of motion, while the continuous phase is given in Eulerian frame of reference.
In contrast to the Euler--Euler method, the Euler--Lagrange method can handle numerous particles with acceptable accuracy while avoiding the enormous computational cost associated with particle-resolved strategies~\cite{Balachandar2010}.

The modeling of the forces acting on and exerted by the particles in the Euler--Lagrange approach is strongly tied to the volume fraction of the particles in the continuous phase~\cite{Elghobashi1994}.
In the most general sense, particle-laden flow can be classified into three categories: very dilute, dilute, and dense suspensions, ordered by increasing volume fraction.
For very dilute flow, the influence of the particulate phase on the continuous phase can be omitted.
Moreover, the mean-free path between the particles is large, such that the probability for inter-particle collisions is negligible for most practical applications.
The resulting interaction in which only the influence of the fluid forces on the particulate phase are considered is called one-way coupling.
For a dilute flow regime, it may be necessary to consider the back scattering of the particulate phase onto the continuous flow field, leading to a two-way coupled approach.
For a further increase in volume fraction, resulting in dense particle-laden flow, the mean distance of the particles decreases such that inter-particle collisions have to be taken into account.
The resulting four-way coupling leads to the highest fidelity Euler--Lagrange approach.
However, the identification of particle collisions comes at extensive computational cost.
Furthermore, for a particle volume fraction approaching unity, the flow is purely collision dominated.
In this case, the point particle method cannot be applied, and more sophisticated computational approaches, such as discrete element methods, must be employed.
In this paper, we restrict ourselves to applications with dense suspensions below the granular flow regime with a strong influence of the interstitial fluid.
Consequently, these applications necessitate the adoption of a four-way coupled approach.
We elect to represent the inter-particle collisions using a deterministic collision model and the hard-sphere approach (binary
collisions) to relax the time step restriction compared to its counterpart, the soft-sphere method, where particle collisions are
resolved in time~\cite{Crowe2011}.

As noted above, high-fidelity simulations of four-way coupled particle-laden flows come at considerable computational cost, much of which is due to the collision operator.
The main bottleneck of particle collision algorithms is the search for pairwise collision partners, generally performed via a
variant of the nearest-neighbour search approach.
The computational effort is linked to the number of eligible collision partners and rises with increasing volume fraction, resulting in numerous collisions per time step and thus a multitude of possible particle pairs.
Hence, several authors proposed optimizations of the original nearest-neighbor search algorithm to efficiently detect potential
particle collision pairs, with comprehensive overviews given by \citet{Sigurgeirsson2001} and \citet{Ching2021}.
These optimized algorithms incorporate approaches to facilitate the early rejection of ineligible particle pairs and may be broadly classified into particle neighbor
lists and cell-based/element neighbor lists approaches.
Particle neighbor lists approaches attach a distance-sorted list of the surrounding particles to each particle in order to discard particle pairs above a given distance threshold.
As these lists contain a considerable amount of redundant information for particles in close proximity, this approach is inherently memory-intensive.
Concurrently, there is no intrinsic mechanism to prompt an update of the particle neighbor lists, which may result in inaccurate omission of particle pairs.
The cell-based neighbor lists approach addresses these deficiencies by first mapping each particle either to its computational mesh
element or to an auxiliary Cartesian grid of bins, also denoted as virtual cell approach or bin neighbor lists~\cite{Norouzi2017,Liu2022}, depending on the cell size chosen.
Both methods have in common that only particles in the same and the adjacent grid cells (node sharing) are considered in the nearest neighbor search.
Cell sizes for the bin neighbor lists are typically chosen equal to the diameter of the largest particle while the virtual cell approach permits arbitrary sizes.
This method has the main drawback that the bins are usually distributed uniformly throughout the domain.
The result is an increased computational overhead for applications with strongly differing and non-uniform element sizes such as
channel/pipe flows or flows around airfoils, as considered in this paper.
This drawback is exacerbated if the particles exhibit velocities across several order of magnitude, rendering this approach
typically memory-intensive and computational expensive~\cite{Ching2021}.
In addition, the optimal bin size for a hard-sphere approach remains an open research question as the use of larger cells incurs
unwarranted additional inspections, while smaller bins unnecessarily limit the time step size~\cite{Sigurgeirsson2001}.
In comparison, the element-based neighbor list method offers the advantage of being inherently suitable for the efficient collision search on meshes with strongly non-uniform and differently sized element shapes.
The omission of a virtual grid implies that in a parallel setting only the computational mesh needs to be efficiently mapped to the
processors, whereas in the bin- and virtual-cell approaches the auxiliary Cartesian background grid also has to be mapped to the
processors.
More recent approaches aim to combine the favorable aspects of the particle neighbor lists and cell-based neighbor lists approaches.
\citet{Yao2004} accelerated the construction of the neighbor list by combining virtual cells with a particle neighbor lists
approach, which, however, can be highly memory-intensive.
\citet{Breuer2012} refined the cell-based neighbor lists and eliminated the necessity to take the surrounding cells into account by
employing two virtual Cartesian grids of different size. This comes at the cost of increased computational overhead, especially in a parallel context.
Subsequent publications by Krijgsman, Ogarko, and Luding~\cite{Ogarko2012,Krijgsman2014} aimed to determine the optimal hierarchical
cell space for multi-level virtual cell-based neighbor list approaches, focusing mainly on soft-sphere methods.
In the context of the bin-based neighbor list approach, an earlier approach
by \citet{Sigurgeirsson2001} resulted in improved efficiency for hard-sphere collisions, but with
the aforementioned drawbacks of the bin-based methods.
A more recent approach was proposed by \citet{Ching2021}, who presented a four-way coupled Euler--Lagrange approach using the
high-order discontinuous Galerkin (DG) method.
The authors aimed to develop an efficient variant of the element-neighbor-list by constructing an element proximity list in reference space.
This list is built after mapping the particle positions from physical space to the unit cube in reference space and subsequent truncation of the number of eligible adjacent elements.
To increase the efficiency, the number of neighboring elements used for the particle pair search is restricted depending on the position of the particle within the element.
This can neglect possible particle collisions, particularly for very distinct particle velocities and trajectories.
Finally, it is worth mentioning that recent works speed up the particle neighbor search by using GPU acceleration, see, e.g., \citet{Liu2022}.
All in all, this renders the element-based neighbor list method (element-binning approach) as the appropriate choice for this work.

The main objective of this paper is to present computationally efficient algorithms for four-way coupled Euler--Lagrange simulations using an element-based neighbor list
approach.
As accuracy and compactness of projection kernels for particle-fluid coupling is extensively covered in literature, see~\cite{Ching2021a} for a more comprehensive overview, we place the focus for particle-fluid coupling on addressing parallel efficiency, especially in the context of high performance computing (HPC).
The proposed inter-particle collision operators builds on this parallelization aspect,
enabling highly efficient and accurate calculation of binary inter-particle collisions in combination with particle-fluid
coupling on arbitrary core counts.
The particular features of this particle collision algorithm compared to the previously presented methods can be summarized as follows.
First, the efficiency and HPC suitability of the proposed algorithm is achieved via the combination of
the particle operator with an MPI+MPI hybrid approach~\cite{Hoefler2013, Zhou2019, Quaranta2021}.
On multinode computations, the novel collision algorithm includes all particles which are in the halo region of its element in the particle collision pair
search without a significant increase in computational work or memory pressure.
Potential load imbalances are handled via a dynamic load balancing procedure.
The proposed algorithm is code-agnostic, which facilitates integration into any Euler--Lagrange framework.
Second, the new inter-particle collision algorithm is implemented in the high-order Euler--Lagrange framework
\elexi\footnote{\url{https://github.com/flexi-framework/elexi}}~\cite{Kopper2023} (so far only two-way coupling).
In \elexi, the carrier phase is discretized by a hybrid discretization operator based on a high-order accurate discontinuous
Galerkin Spectral Element Method (DGSEM) and a localized low-order finite volume operator, while a Lagrangian approach is employed for the discrete phase.
As such, the proposed algorithm ties into the existing capabilities of \elexi of handling complex geometries on unstructured grids featuring boundary conditions, possibly curved elements, and hanging nodes.
In combination, these features enable an efficient and highly accurate treatment of inter-particle collisions in a compressible
carrier phase on arbitrary core counts which is demonstrated by its excellent scaling properties and efficient memory utilization on massively parallelized systems.
Finally, the \elexi framework is to the author's knowledge the first open-source solver for one- to four-way coupled compressible particle-laden flows in complex geometries using the high-order DG method.

This primary focus of this work is on the implementation, the parallelization challenges and the application of discrete particles in dense
suspension within a continuous compressible flow field.
In~\cref{sec:theory}, the underlying equations for both the continuous and discrete phase are presented, including the fluid-particle coupling,
particle-wall interactions, and inter-particle collisions.
This is followed by a brief outline of the numerical treatment of these equations in~\cref{sec:methods}.
The parallelization strategies for the collision and projection operator are detailed in~\cref{sec:implementation} and validated in~\cref{sec:validation}, followed by a demonstration of the parallel performance in~\cref{sec:parallel}.
The capabilities of our four-way coupled Euler--Lagrange approach are demonstrated in~\cref{sec:application} using two
challenging real-world applications.
This paper closes with a brief discussion in~\cref{sec:conclusion}.

\section{Theory}%
\label{sec:theory}

\subsection{Continuous Phase}%
\label{sec:theory:nse}
The fluid field is governed by the compressible unsteady Navier--Stokes--Fourier equations, given in vectorial form as
\begin{align}
  \frac{d\cons}{dt} + \nabla\cdot \fphys\left(\cons,\nabla\cons\right) = \source,
  \label{eq:theory:fluid:NSE}
\end{align}
where $\cons = \smash{[\rho,\rho u_1,\rho u_2,\rho u_3,\rho e]^T}$ is the vector comprising the conservative variables with
$\rho$ as the fluid density, $u_i$ the $i$-th component of the velocity vector and $e$ the total energy per unit mass.
The source term $\source$ accounts for the influence of the dispersed phase on the fluid in two- or four-way coupled regimes.
The physical flux $\fphys$ is composed of the inviscid Euler and the viscous fluxes.
The equation system is closed by the
equation of state of a calorically perfect gas. The dynamic viscosity $\dynvisc$ is obtained from Sutherland's
law~\cite{Sutherland1893}, while the heat flux is given by Fourier's law.
Following Stokes' hypothesis, the bulk viscosity is set to zero.

\subsection{Dispersed Phase}%
\label{sec:theory:maxey}

According to the Lagrangian point particle approach, particles are treated as discrete points with mass $\partmass$ and diameter
$\partdiam$ which move in a Lagrangian manner
according to the following system of ordinary differential equations (ODEs)
\begin{align}
  \frac{d \partpos}{dt} &= \partvel, \\
  \partmass \frac{d \partvel}{dt} &= \mathbf{F} = 3\pi\dynvisc \partdiam \dragfactor \left(\fluidvel - \partvel \right) +
  \forcelift, \label{eq:mrg:mom} \\
  \moinertia \frac{d \vorticity_p}{dt} &= \torque = \fluiddens \frac{\partdiam^5}{64} C_w \angularvel \abs{\angularvel}, \\
  \partmass c_p \frac{d T_p}{dt} &= Q_p = \pi \partdiam \kappa_c (T_f - T_p) \mathrm{Nu}, \label{eq:mrg:T}
\end{align}
with the particle position in physical space $\smash{\partpos=[\partpos[][1], \partpos[][2], \partpos[][3]]^T}$ and the particle velocity
$\partvel$ obtained from the integration of the first two ODEs.
\Cref{eq:mrg:mom} is approximated by the Maxey--Riley--Gatignol (MRG) equation~\cite{Maxey1983,Gatignol1983} with the empirical
correction for higher particle Reynolds numbers and the drag factor $\dragfactor$, computed according to \citet{Loth2021}.
The angular particle velocity is $\smash{\vorticity_p=\nabla \times \partvel}$, $\moinertia = \frac{\pi}{60} \partdens \partdiam^5$
is the moment of inertia of a spherical particle, $\smash{\angularvel=\frac{1}{2}(\nabla \times \fluidvel) - \vorticity_p}$ is the relative
fluid-particle angular velocity, $\torque$ is the
torque and $C_w$ is a correction factor for higher Reynolds numbers proposed by \citet{Dennis1980}.
\Cref{eq:mrg:T} describes the change of the particle temperature $T_p$, where $Q_p$ is the heat transfer term, $\mathrm{Nu}$ the
Nusselt number, $c_p$ the specific heat of the particle, and $\kappa_c$ the thermal conductivity of the continuous phase.

\subsection{Fluid-Particle Coupling}%
\label{sec:theory:two_way}

For two- and four-way coupled flow, the influence of the particulate on the continuous phase is modeled using the particle-source-in-cell approach proposed by
\citet{Crowe1977}. In this approach, the forces acting on the particles and the corresponding work appear as a source term, $\source =
[0,\sourcei_{m,1},\sourcei_{m,2},\sourcei_{m,3},\sourcei_e]$ in the
momentum equations and the energy equation, respectively.
The source terms for the momentum equations $\source_m=[\sourcei_{m,1},\sourcei_{m,2},\sourcei_{m,3}]$ and the energy equation
$\sourcei_e$ at a point $\vvec{x}_{ijk}, \ i,j,k \in \mathbb{N}_{>0}$ are given by
\begin{align}
  \source_m &= - \mathcal{P}\left\{\left(\mathbf{F} \right), \vvec{x}_{ijk} \right\},\\
  \sourcei_e &= - \mathcal{P}\Big\{\source_m \cdot \partvel + Q_p, \vvec{x}_{ijk}\Big\},
  \label{eq:theory:source}
\end{align}
with the projection operator $\mathcal{P}\{\cdot,\cdot\}$, which projects the source term onto the grid.
As such, this approach is also referred to as particle deposition.
Within this work, the influence of the source term is linearly imposed onto the corner nodes of the host element,
as a $C_0$-continuous version of the inverse distance weighting (IDW) interpolation~\cite{Shepard1968, Stindl2015}.
Given the eight corner nodes of a hexahedral elements $\{\boldsymbol{\xi}_n | \boldsymbol{\xi} \in [-1,1]^3\}_{n=1}^{8}$ and the
degrees of freedom (DOF) $\{\boldsymbol{\xi}_{ijk}\}_{i,j,k=1}^{\ppn}$, the IDW interpolation function for an arbitrary
variable $a\in\mathbb{R}$ and a single particle results in
\begin{align}
  \mathcal{P}\left\{ a, \vvec{x}_{ijk} \right\} = \frac{J_{ijk} w_\xi(\boldsymbol{\xi}_{ijk})
  w_\xi(\boldsymbol{\partrefpos}) \ a}{\sum_{n=1}^{8} J_{n}^{1} w_{n}^{1}}
  \label{eq:theory:deposition}
\end{align}
with $w_\xi(\boldsymbol{\xi}) = \sum_{n=1}^8 \prod_{d=1}^3 (\xi_{n}^d-\xi^d)/2$ and the particle position in reference space
$\partrefpos\in[-1,1]^3$.
Here, $w(\boldsymbol{\xi}_n)$ are the weights defined at the corner node $n$ and $J_n$ is the corresponding Jacobian such that
the product of both denotes the volume spanned by node $n$.
Different projection operators such as a Dirac delta function are available in the code should the user prefer another choice for the projection operator $\mathcal{P}$.

\subsection{Inter-Particle and Particle-Wall Interactions}

In the following, we briefly describe the physical model for inter-particle and particle-wall collisions.

\subsubsection{Inter-Particle Collisions}%
\label{sec:theory:four_way}

In the hard-sphere approach, only binary collisions between particles are considered and particle deformations are neglected.
The jump relations describing the change in momentum are given by \citet{Crowe2011} as
\begin{align}
  \partmass[1] (\partvel[1]-\partvel[1]^{(0)}) = &  \mathbf{J}, \\
  \partmass[2] (\partvel[2]-\partvel[2]^{(0)}) = & -\mathbf{J}, \\
  I_{p,1} (\vorticity_{p,1}-\vorticity_{p,1}^{(0)}) = & \partdiam[1]/2 (\normalvec_p \cross \mathbf{J}), \\
  I_{p,2} (\vorticity_{p,2}-\vorticity_{p,2}^{(0)}) = & \partdiam[2]/2 (\normalvec_p \cross \mathbf{J}),
\end{align}
where $\mathbf{J}$ is the unknown impulsive force, and $\normalvec_p$ the unit vector from particle $1$ to particle $2$. Variables with the
superscript $(\cdot)^0$ are pre-collision quantities, while no superscript denotes the unknown post-collision quantities.
Following~\cite{Crowe2011}, the impulsive force, given as
\begin{align}
  \mathbf{J} &= J_n \normalvec_p + J_t \mathbf{t}_p, \\
  J_n &= - \partmass[r] (1+\cor{n}) (\partvel[r]^{(0)} \cdot \normalvec_p) < 0, \\
  J_t &= \frictioncoeff J_n < 0,
\end{align}
is computed based on the relative particle motion, $\partvel[r] = \partvel[1]-\partvel[2]$, the relative mass, $\partmass[r] =
\frac{\partmass[1]\partmass[2]}{\partmass[1]+\partmass[2]}$, and two parameters,
the normal coefficient of restitution $\cor{n}$ and Coulomb's law of friction with the friction coefficient $\frictioncoeff$.
The tangential component of the relative velocity of the contact point,
\begin{align}
  \partvel[r,t]     &= \partvel[c]^{(0)} - (\partvel[c]^{(0)} \cdot \normalvec_p) \normalvec_p, \\
  \partvel[c]^{(0)} &= \partvel[r]^{(0)} + r_1 \vorticity_{p,1}^{(0)} \cross \normalvec_p + r_2 \vorticity_{p,2}^{(0)} \cross \normalvec_p,
\end{align}
determines if the particles are sliding (first condition) or
sticking (second condition) which then leads to
\begin{align}
  J_t = \text{max} \left(\frictioncoeff J_n, -\frac{2}{7} (1+\cor{n}) \partmass[r] \abs{\partvel[r,t]^{(0)}} \right)
\end{align}
and the tangential vector given as $\mathbf{t}_p=\partvel[r,t]^{(0)} \abs{\partvel[r,t]^{(0)}}^{-1}$.

\subsubsection{Intersection of Particles with Solid Walls}%
\label{sec:theory:rebound}
Particle-wall collisions are also modeled via the hard-sphere approach, i.e., they are not resolved in time but handled in an a
posteriori manner. Thus, the time step $\dtstage$ of the current Runge-Kutta stage has to be greater than the time a particle requires to collide with a wall
($\dtstage > \dtcoll$).
The change in particle momentum and angular particle momentum are modeled according to \citet{Crowe2011} as
\begin{align}
  \partmass[2] (\partvel[2]  + 2 (\partvel[1] \cdot \normalvec) \normalvec)- \partmass[1] \partvel[1] = \partmomentum, \\
  \moinertia_2 \vorticity_2 - \moinertia_1 \vorticity_1 = - \frac{\partdiam[1]}{2} \left[\normalvec \times (\partmass[2] \partvel[2]
  - \partmass[1]\partvel[1])\right],
\end{align}
where $(\cdot)_2$ are quantities after the impact and $(\cdot)_1$ before it.
The change in momentum is designated as $\partmomentum$, and $\normalvec$ is the normal vector of the boundary.
For a perfectly reflective wall, i.e., an elastic collision, $\partmomentum=0$, while for a plastic deformation
$\partmomentum\neq0$.
Here, only the former is considered, the reader is referred to, e.g.,~\cite{Bons2017, Schwarz2022} for further
details on plastic particle-wall collisions.

\section{Numerical Methods}%
\label{sec:methods}
In the following, we briefly discuss the numerical treatment of the governing equations for the fluid and dispersed phase.

\subsection{Discontinuous Galerkin Spectral Element Method}%
\label{sec:methods:dgsem}
The Navier-Stokes-Fourier equations are solved via the Discontinuous Galerkin Spectral Element Method (DGSEM), where the
computational domain $\smash{\Omega \subseteq \mathbb{R}^3}$ is discretized by non-overlapping, (non-)conforming hexahedral elements with
six possibly curved element faces. %
Subsequently, the considered governing equations are transformed into the reference coordinate system $\smash{\boldsymbol{\xi} = [\xi_1,
\xi_2, \xi_3]^T}$ of the reference element $E=[-1,1]^3$ via the mapping $\smash{\mathbf{x} = \boldsymbol{\chi}(\boldsymbol{\xi},t)}$,
$\mathbf{x} \in \Omega$.
The weak form is retrieved by a discrete $L_2$ projection of the governing equation in reference space onto the test space composed of polynomials $\testfunc(\boldsymbol{\xi})$ up to degree $\ppn$,
followed by an application of Gauss' theorem, yielding
\begin{align}
  \int_E J \frac{\partial \cons_h}{\partial t} \testfunc(\boldsymbol{\xi}) \,d\boldsymbol{\xi}
  +& \int_{\partial E} \fnumref \testfunc(\boldsymbol{\xi}) \,d\vvec{S} \nonumber \\
  -& \int_E \fphysref(\cons_h, \nabla \cons_h) \cdot \nabla_\xi \testfunc(\boldsymbol{\xi}) \,d\boldsymbol{\xi} = 0,
  \label{eq:dgsem}
\end{align}
with the Jacobian $J$ of the mapping, the outward pointing normal vector $\normalvec$, the contravariant flux vector $\fphysref$ and the numerical flux normal to the element face $\fnumref$.
To obtain an efficient discretization, the element-local solution $\cons_h = \cons_h(\boldsymbol{\xi},t)$ is approximated by a tensor product of one-dimensional nodal
Lagrange basis functions $l$ of degree $\ppn$
\begin{align}
  \cons_h(\boldsymbol{\xi},t) = \sum_{i,j,k=0}^{\ppn} \consdofs_{ijk}(t) l_i(\xi^1)l_j(\xi^2)l_k(\xi^3),%
  \label{eq:dgsem_interpolation}
\end{align}
with the nodal degrees of freedom $\consdofs_{ijk}(t)$.
For the numerical integration of \cref{eq:dgsem}, the Legendre-Gauss quadrature with $(\ppn+1)^3$ Legendre-Gauss points is employed.
If not stated otherwise, the numerical flux is approximated via the approximate Riemann solver by \citet{Roe1981},%
with the entropy fix by \citet{Harten1983}.
The viscous fluxes are computed with the BR1 scheme~\cite{Bassi1997}.
To mitigate aliasing errors, the flux is split according to \citet{Pirozzoli2011}.
The shock capturing procedure is based on a second-order accurate finite volume (FV) (subcell) scheme with $(\ppn+1)^3$ integral means per DG element~\cite{Sonntag2017}.
Following the method of lines approach, the explicit low-storage fourth-order accurate Runge--Kutta (RK) scheme by \citet{Carpenter1994} is employed for the integration in time.
The reader is referred to~\cite{Sonntag2017, Beck2014, Hindenlang2012, Krais2019} for further details on DGSEM and applications.
The methods presented in this work are implemented in the open-source framework~\elexi.

\subsection{Particle Localization and Time Integration}%
\label{sec:methods:tracking}
In contrast to the aforementioned fluid phase, the dispersed particulate phase is tracked in physical space.
While tracking approaches in reference space are reported to be more robust due to their natural localization within each element, these methods often do not inherently consider boundary conditions~\cite{Ortwein2019}.
\elexi elects to follow the approach described in \citet{Ortwein2019} by tracking particle-face intersections in physical space following methods known from ray tracing in computer graphics.
Neglecting inter-particle collision, the particle motion is thus computed using the following four steps.

\subsubsection{Emission and Localization}
Particles are emitted in parallel at physical locations determined from predefined spatial distribution functions.
The particle host element is located with the help of a Cartesian background mesh where each background mesh element contains a mapping to each overlapping computational mesh element.
For each overlapping element, the location of the particle in reference space is determined by finding the root of
\begin{equation}
  \partpos - \boldsymbol{\chi}(\partrefpos) = 0
\end{equation}
via Newton's method.
The particle host element is found once $\partrefpos$ satisfies $\partrefpos \in [-1, 1]^3$.

\subsubsection{Field Evaluation}
The fluid field at the particle's center of mass is calculated from a straightforward evaluation of the DG polynomials, thus
\begin{equation}
  \cons_h(\partrefpos,t) = \sum_{i,j,k=0}^\ppn \hat{\cons}_{ijk}(t) l_i(\partrefpos^1) l_j(\partrefpos^2) l_k(\partrefpos^3).
\end{equation}
The particle interpolation reduces to a linear interpolation of the conserved variables to the particle position for the
second-order FV subcells scheme employed for shock capturing, see~\cref{sec:methods:dgsem}.

\subsubsection{Time Integration}
Particles are integrated in time using linear segments such that the particle path within a single Runge--Kutta (RK) stage, i.e., $t \in [t^{n},t^{n+1}]$, is described by
\begin{align}
  \partpos(t; \alpha) &= \partpos(t^{n}) + \alpha \frac{\parttrajectory}{\vert \parttrajectory \vert},& \alpha &\in [0, \vert
  \parttrajectory \vert],\label{eq:methods:path}\\
  \parttrajectory     &= \partpos(t^{n+1}) - \partpos(t^{n}).\label{eq:methods:trajectory}
\end{align}
Here, $\parttrajectory$ describes the linear path segment while $\alpha$ is the relative displacement along $\parttrajectory$.

\subsubsection{Face Intersections}
The linear path segments from \cref{eq:methods:path,eq:methods:trajectory} need to be checked for face intersections to determine updates to the particle host element and/or applications of boundary conditions.
For curvilinear sides, the intersection location is performed using a dimension reduction approach based on Bézier clipping following the de Casteljau subdivision approach.
Each element side is mapped onto a Bézier polynomial surface while preserving the shape of the element side, yielding a representation as
\begin{align}
  \bezierpoly(\xi,\eta) = \sum_{m=0}^{\ppn_{\text{\textit{geo}}}} \sum_{n=0}^{\ppn_{\text{\textit{geo}}}} \bezierdofs_{mn} \bezierbasis_m(\xi) \bezierbasis_n(\eta).
\end{align}
Here, $\ppn_{\text{\textit{geo}}}$ describes the polynomial degree of the mapping, $\bezierdofs$ are the Bézier control points and $(\xi,\eta) \in [-1,1]^2$ the side coordinates in reference space.
With this formulation, a face intersection is found if and only if the root of
\begin{align}
  \partpos(t,a) = \partpos(t^{n}) + \alpha \frac{\parttrajectory}{\vert \parttrajectory \vert} \overset{!}{=} \bezierpoly(\xi,\eta) \;\ni\! \begin{array}{l} \makebox[\widthof{$(\xi,\eta)$}][l]{t} \in (t^{n},t^{n+1}),\\ (\xi,\eta) \in [-1,1]^2\end{array}
\end{align}
exists.
Alternatively, simpler approaches are applied if a face is detected to collapse to a linear or bi-linear side.
For more details on the particle tracking approach, see \citet{Ortwein2019}.

\subsection{Particle Collisions}%
\label{sec:methods:collision}
To check for the possibility of particle collision between two consecutive Runge--Kutta stages $t^{n}$ and $t^{n+1}$, an a
posteriori approach is chosen. Here, particle collisions are checked after the particles moved from $t^{n}$ to $t^{n+1}$. Hence, it
is again assumed that the particle movement in one RK stage is according to a straight line.
Two particles collide if there exists a time span $\dtcoll > 0$ for which  $\dtcoll \leq \dtstage \in [0; t^{n+1}]$, i.e., if the
following quadratic equation
\begin{align}
  \abs{\Delta x_p + \dtcoll \Delta v_p}^2 &= \frac{(\partdiam[1]+\partdiam[2])^2}{4},\label{eq:methods:collision:detect1} \\
  \Delta x_p = \dxpart, \ \Delta v_p &= \dvpart
  \label{eq:methods:collision:detect2}
\end{align}
has at least one solution for which $\dtcoll \leq \dtstage$.
If there exist two positive solutions of \cref{eq:methods:collision:detect1,eq:methods:collision:detect2} and the lower $\dtcoll$ fulfills the condition $\dtcoll
\leq \dtstage$, then the collision time is given as $\tcoll=t^{n}+\dtcoll$, otherwise no collision occurs in the current RK stage.
Finally, the collision is only considered valid if the following two conditions hold:
First, the collision occurred before an intersection with a domain boundary. Second, the particles approach each other.

After collision pairs are detected, the particle velocities $\partvel(t^{coll})$ after the collision are determined as described
in~\cref{sec:theory:four_way} and the particle positions are updated accordingly, thus
\begin{align}
  \partvel[1](t^{coll}) &=  \frac{\mathbf{J}}{\partmass[1]}+\partvel[1]^{(0)},\label{eq:methods:collision:update_vel1} \\
  \partvel[2](t^{coll}) &= -\frac{\mathbf{J}}{\partmass[2]}+\partvel[2]^{(0)},\label{eq:methods:collision:update_vel2} \\
  \partpos(t^{coll})    &=  \partpos^{(0)} + (\dtstage - \dtcoll) \partvel(t^{coll}). \label{eq:methods:collision:update_rho}
\end{align}

\section{Implementation Details}%
\label{sec:implementation}

The following chapter details the implementation choices for the code parallelization of the four-way coupled framework detailing
the most critical parts for parallel performance, the particle deposition and collision operators. It concludes with a brief discussion of the load balancing strategy. For further details the reader is referred to \citet{Kopper2023}.

\subsection{Parallelization}%
\label{sec:implementation:parallelization}
\elexi is parallelized using pure Message Passing Interface (MPI) communication~\cite{mpi40}. Mesh elements are pre-sorted along a space-filling curve (SFC) by the open-source mesh generator High-Order PreProcessor HOPR\footnote{\url{https://github.com/hopr-framework/hopr}}~\cite{Hindenlang2015}.
Solution data for the continuous phase is structured to form linear memory segments for communication and is stored using the distributed memory paradigm.
As the computational stencil of the DG method used for the continuous phase is element-local, solely the numerical flux is exchanged between individual elements using non-blocking communication.
At the same time, intra-element computation is utilized to enable efficient latency hiding.

As the discrete phase is allowed to move arbitrarily between the mesh elements, the processor-local geometry information is enriched by the spatially surrounding elements to create a halo region.
An efficient communication-free two-step search algorithm determines the eligible elements~\cite{Kopper2023}.
This halo region is stored on each individual compute node using the MPI-3 shared memory (SHM) paradigm.
Thus, each processor has direct access to the complete mesh geometry for particle tracking and can determine the final particle position while incorporating boundary conditions.
Only once the final particle position is known, the particles are sent to the new host processor using non-blocking MPI communication.
\elexi attempts to finish the particle tracking as early as possible to hide the communication latency behind the local computational load from the continuous phase.
\begin{figure}[!tb]
  \includegraphics[width=\columnwidth]{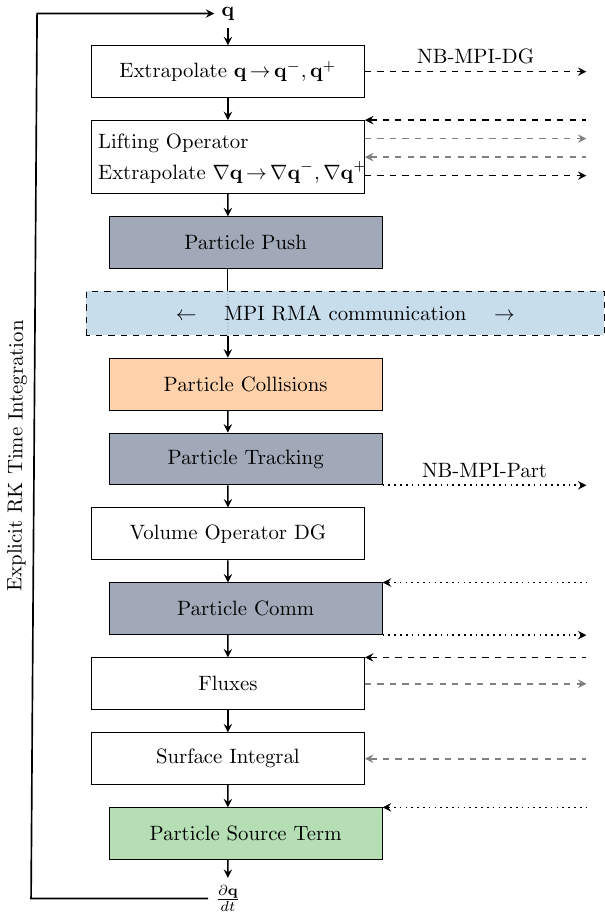}
  \caption{Flow chart of the discontinuous Galerkin operator for 4-way coupled particle-laden flow. Dashed gray line indicate
  non-blocking MPI communication for the DG operator (NB-MPI-DG), dotted black lines MPI communication for the particle operator
(NB-MPI-Part). DG operations are shaded in white. Particle operations are shaded in gray, with the particle collision operator highlighted in orange and the
particle deposition operator in green. See \citet{Kopper2023} for a detailed breakdown. The blue block represents MPI RMA operations introduced for 4-way coupling.}%
  \label{fig:implementation:collision:latency}
\end{figure}

\subsection{Deposition}%
\label{sec:implementation:deposition}
If the particle volume fraction exceeds the limit of dilute dispersed flow, the effect of the dispersed particles on the continuous fluid must be considered.
This is achieved through deposition of the forces acting on the particles as a corresponding work on the continuous phase with the
sign reversed, thus appearing as source terms in the Navier--Stokes--Fourier equations, see~\cref{eq:theory:fluid:NSE}.
As the particle positions generally does not coincide with a fluid DOF, \elexi features several mechanisms to conservatively project the source terms onto the fluid solution.
The most straightforward approach assigns the complete source term to the nearest DOF (Dirac delta function).
While this results in a highly localized deposition, sharply non-uniform particle concentrations can cause strong oscillations and nonphysical solutions in the DG polynomials.
Smoothing approaches based on shape functions can alleviate this issue by extending the influence across multiple solutions points
but require large communication stencils and are slightly less accurate~\cite{Stindl2015}.
As a consequence, the most efficient approach available in \elexi implements a linear deposition approach based on IDW
interpolation, see~\cref{eq:theory:deposition} where the deposition radius is chosen identical to the cell size.
Here, each element corner node is assigned a unique node identifier during pre-processing with HOPR while taking periodic boundary conditions into account.
Since the high-order DG method utilizes a low number of elements while preserving high numerical accuracy through the arbitrary
order polynomial basis, each compute node can allocate an MPI-3 SHM array sufficient to store the deposition source terms for the element corners nodes of the complete computational domain.
Deposition on the corner nodes is performed locally on each compute using atomic \verb|MPI_ACCUMULATE| calls.
The source terms are summed up across the compute nodes using non-blocking \verb|MPI_IALLREDUCE| operations once the compute node-local deposition is complete.
This procedure results in a conservative deposition approach which is numerically stable, computationally efficient, and features reasonable computational accuracy.

\subsection{Collision Operator}%
\begin{figure}[!tb]
  \includegraphics[width=\columnwidth]{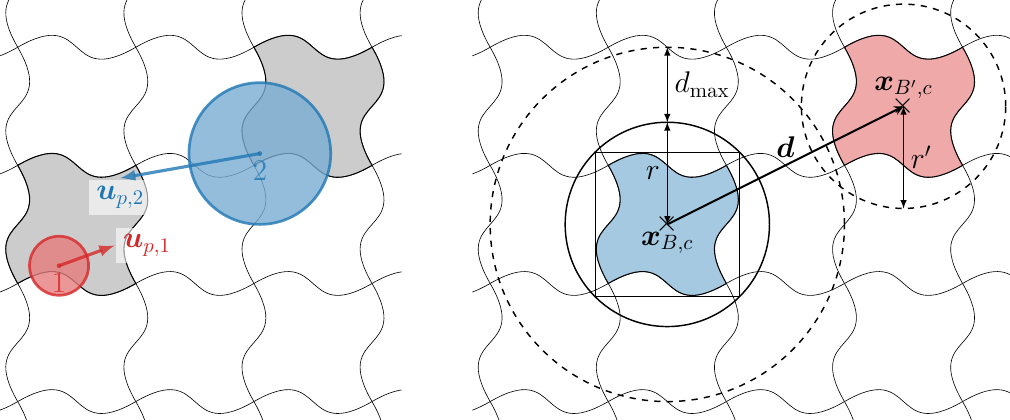}
  \caption{Sketch of halo region determination for particle collisions. Left: Theoretical extent of the Lagrangian particles and their host elements (shaded gray). Right: Spheres of influence for possible particle-particle collisions for the same elements.}%
  \label{fig:implementation:collision:halo}
\end{figure}

\label{sec:implementation:collision}
The distinct feature of the proposed collision operator is the combination of an element-based binning method with an MPI+MPI hybrid approach.
The computation of the collision operator is the most expensive computational operation.
As \elexi calculates exact (binary) hard-sphere collisions, this requires comparisons of all particle trajectories within a given sphere around each particle position.
An early reduction of the amount of potential collision partners which need to be checked is thus crucial for acceptable code performance.
At the same time, accurate results require that no eligible particles are omitted, which poses a challenge during parallel runs.
Defining the barycenters of two elements as $\boldsymbol x_{B,c}$ and $\boldsymbol x_{B',c}$ with the corresponding convex hull radii $r$ and
$r'$, respectively, a particle needs to be considered if its host element is within
\begin{align}
\vert \boldsymbol x_{B',c} - \boldsymbol x_{B,c} \vert = \vert \boldsymbol d \vert \leq d_\mathrm{max} + r + r',
\end{align}
see \cref{fig:implementation:collision:halo}.
Here, $d_\mathrm{max}$ is the maximum distance a particle can travel during one time step and can be calculated from the velocity of the
fastest overall particle and the time increment of the explicit Runge--Kutta scheme.

Although element-based binning reduces memory requirements, retaining complete particle data on a per-task basis nevertheless results in excessive memory pressure.
In order to reduce memory pressure, intranode information is unique, while a ghost layer, or halo region, is created to avoid blocking on inter-node information exchange.
This halo region contains for each mesh element the list of other elements with potential collision partners, which are chosen
depending on the maximum possible distance a particle can travel in one computational time step (here RK stage).
\elexi utilizes its MPI-3 shared memory paradigm to perform this element-based binning approach while simultaneously minimizing the amount of particle data which needs to be exchanged.
During code initialization, an element-based mapping array based on the physical distance is constructed on each compute node, comprising for each mesh element the list of other elements close enough for potential collision partners.
Periodic boundaries are incorporated through virtual element shifts during the distance calculation.
Each list is sorted according to the global element index to ensure determinism.
As the mapping is build on the global mesh information, the list automatically contains elements inside the halo region which are residing on other threads.
The union of the individual lists contain all elements required for particle collisions on a given compute node.

\begin{figure}[!tb]
  \includegraphics[width=\columnwidth]{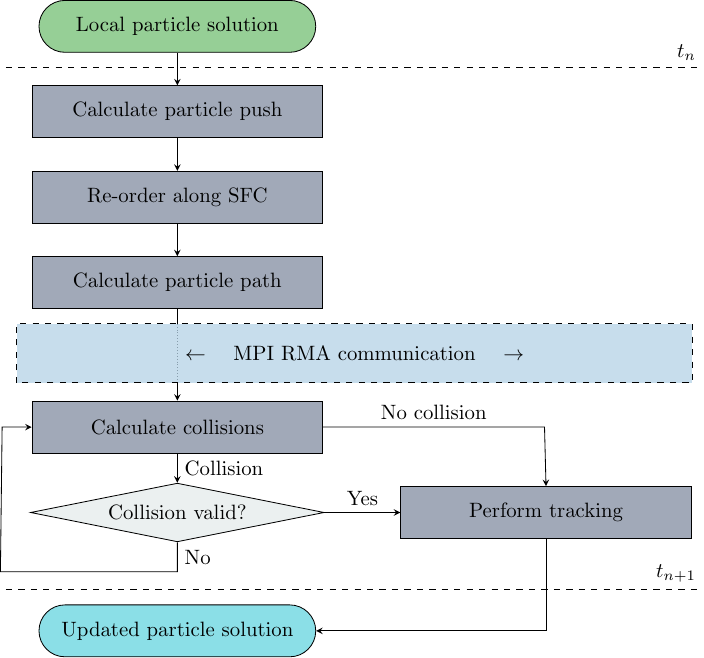}
  \caption{Algorithm for parallel exact hard-sphere collision detection.}%
  \label{fig:implementation:collision:algorithm}
\end{figure}

\begin{figure}[!tb]
  \includegraphics[width=\columnwidth]{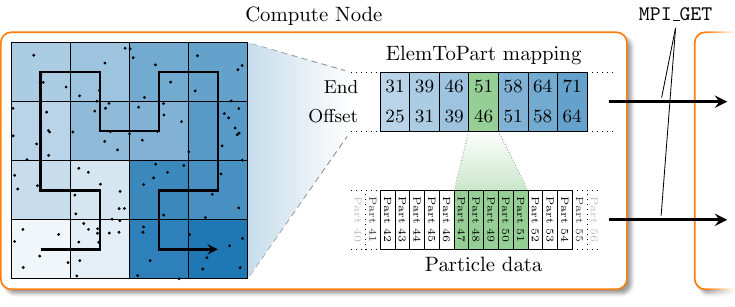}
  \caption{Data structure and inter-node communication for the element-particle mapping and the actual particle data.}%
  \label{fig:implementation:collision:shared}
\end{figure}

During code execution, prior to entering a particle collision operator step, the particles are sorted according to their host element along the space-filling curve using a linked-list approach.
From this sorting, both the number of particles in each element and their global index is obtained.
This total number of particles on each compute node is broadcast using an \verb!MPI_EXSCAN! operation while simultaneously starting an MPI Remote Memory Access (RMA) epoch on the particle lists and data.
\Cref{fig:implementation:collision:algorithm} illustrates the positioning of these operations within the time-stepping algorithm.
The communication overhead is hidden behind the calculation of the remaining particle paths which iterates until each particle encounters a
computational boundary.
Once the communication finishes and the MPI call returns, each process can obtain the required particle data.
\Cref{fig:implementation:collision:shared} gives an overview of the performed memory operations and communication.
Each compute node fetches the indices of particles inside eligible elements using non-blocking \verb!MPI_GET! calls.
With this information, the position of the particle data on each thread can be computed which is again fetched into compute node-local shared memory using non-blocking RMA operations.
The data is enriched with the time of flight for each particle, measured from its current position until it reaches its next boundary intersection.
This information is required for the collision step as collisions occurring outside the computational domain are invalid
solutions.
Once these operations are finished, the position and trajectories of each particle within the halo distance around a compute node is stored in MPI-3 shared memory on each compute node.

The actual collision detection is performed in parallel on each processor.
Every thread loops over its local elements to obtain the list of elements containing possible collision partners.
As this list contains elements in the halo region and is uniquely sorted along the space-filling curve, each thread is ensured to
detect collisions in deterministic order, thus ensuring kinetic energy preservation.
Each particle pair is checked for exact hard-sphere collisions using \cref{eq:methods:collision:detect1,eq:methods:collision:detect2}.
The obtained time of collision is checked against the time of flight until encountering a boundary condition to eliminate physically incompatible solutions.
Once a valid collision solution is obtained, the particles are moved to the collision location and their momentum is updated using
\cref{eq:methods:collision:update_vel1,eq:methods:collision:update_vel2,eq:methods:collision:update_rho}.
Since only one collision is permitted per particle within each time step due to the hard-sphere approach, particles with confirmed collisions are removed from the list of eligible collision partners and transferred to the main tracking algorithm to determine their final location and host element.

\subsection{Load Balance}%
\label{sec:implementation:loadbalance}
While the continuous Euler phase exerts a fixed computational load per grid element, the computational effort of the dispersed Lagrangian phase is dependent on the particle positions and only weakly correlated with the underlying grid.
As the particle distribution generally cannot be computed a priori, \elexi relies on dynamic load balancing to alleviate most of the
disparities in computational load~\cite{Kopper2023}.
Each stage of the time stepping algorithm in \cref{fig:implementation:collision:latency} is equipped with conditional high-precision
timers to compute the total time spent in each stage.
Prior to triggering the actual load balancing, these timers are activated together with counters for the particle number and collision partners in each cell.
After initiating load balancing, the recorded computation time is distributed among the cells in proportion to their contribution to the total particle processing time. This contribution is determined using the relative fractions derived from the particle and collision counters for each cell.

\section{Validation}%
\label{sec:validation}
Before turning to the application cases, we validate the various building blocks for particle-laden flow, from the projection
function in the particle source term to the particle collision operator.
The reader is referred to \citet{Kopper2023} for details on the validation of the particulate phase such as convergence properties
and the correctness of the one- and two-way coupling.
An extensive validation of the continuous phase is given in \citet{Krais2019}.
In the following, only viscous effects such as drag and heating are considered for the particulate phase.
Moreover, particle collisions are assumed to be purely elastic, i.e., $\frictioncoeff=0$ and $\cor{n}=1$, such that the tangential
component of the impulse is zero and the normal component reduces to $\smash{J_n = - 2 \partmass[r] (\partvel[r]^{(0)} \cdot \normalvec_p)}$.
Both phases are integrated in time by a fourth-order accurate explicit Runge-Kutta scheme~\cite{Carpenter1994} with
a relative Courant--Friedrichs--Lewy (CFL) number of $\text{CFL}=0.9$. The Riemann solver by Harten-Lax-Van-Leer-Einfeldt (HLLE)~\cite{Einfeldt1991} is employed for the approximation of the numerical flux function.
The high-order DGSEM is used for the spatial discretization of the continuous phase, which is assumed to be inviscid, i.e., only the Euler
equations are considered in this section, if not stated otherwise.
It has to be noted that all quantities are non-dimensional.

\subsection{Particle Deposition}%
\begin{figure}[!tb]
  \includestandalone[width=\columnwidth]{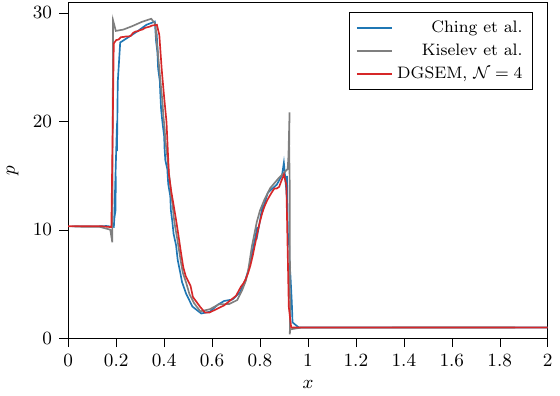}
  \caption{Pressure distribution along the centerline for the interaction of a shock wave with a spanwise-inhomogeneous
    particle cloud at a non-dimensional time of $t=0.24$. Results are compared to the numerical results by~\citet{Kiselev2006} (third-order finite difference (FD) method)
  and~\citet{Ching2021a} (DG, $\ppn=3$) at $t=0.75\si{m\second}$.}%
  \label{fig:validation:partdepo_pressure}
\end{figure}

\begin{figure*}[!tb]
  \centering%
  \includestandalone[width=\linewidth]{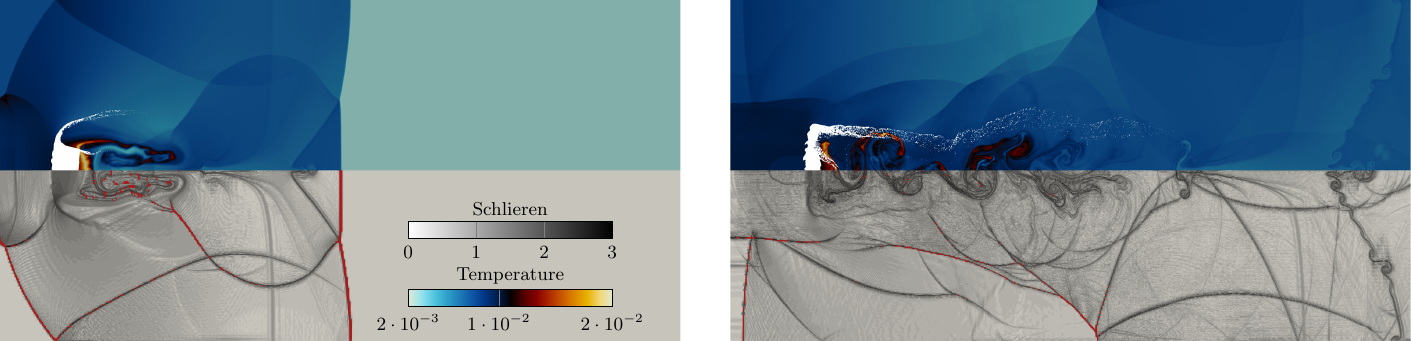}
  \caption{Top: Temperature distribution for the interaction of a shock wave with a spanwise-inhomogeneous
    particle cloud at non-dimensional times $t=0.8$ (left) and $t=1.86$ (right) with particles. Bottom: Pseudo-Schlieren
    ($\abs{\nabla \rho}$) profile with the FV subcell
  distribution denoting the shock capturing highlighted in red.}%
  \label{fig:validation:partdepo_density}
\end{figure*}

First, the validity of the employed particle deposition procedure is demonstrated using the interaction of a shock wave with a spanwise-inhomogeneous
particle cloud, following~\cite{Kiselev2006,Jacobs2012,Ching2021a}. Thus, only a two-way coupling is considered.
Since \elexi is a pure three-dimensional code, a quasi two-dimensional setting is chosen where the computational domain $\Omega=[0,5.5] \times [-1.1,1.1] \times [0, 0.001]$ is discretized by $500 \times 250 \times 1$ elements with
$\ppn=4$.
The initial solution is a right moving shock wave with a Mach number of $M=3$, located at $x=0.315$, and a pre-shock state of
$(\rho,\fluidvel,p)=(1.2,\mathbf{0},1)$.
Non-reflecting boundary conditions are prescribed at the left and right boundary conditions. Symmetry boundary conditions are
utilized at the upper and lower boundaries, while periodic conditions are employed in the remaining direction.
A cubic particle cloud with $\partdiam=\num[scientific-notation=true,retain-unity-mantissa=false]{1.e-4}$ and $\partdens=\num{8900}$ is initialized in the rectangular region
$[0.315,0.634]\times[-0.16,0.16]\times[0,0.001]$ with a volume fraction of around $4\%$, resulting in a total number of
\num{5.e5} computational particles. Each computational particle represents \num{16} real particles.
A fluid with a constant dynamic viscosity of $\mu=\num{1.7144e-08}$ is assumed.
As illustrated in~\cref{fig:validation:partdepo_pressure}, the resulting centerline pressure profiles are comparable to~\cite{Kiselev2006,Ching2021a}.
Slices of the corresponding temperature and pseudo-Schlieren ($\abs{\nabla \rho}$) profile at non-dimensional simulation times $t=0.8,1.86$ are depicted in~\cref{fig:validation:partdepo_density}.

\subsection{Simple Particle Collision}
\begin{figure}[!tb]
  \includestandalone[width=0.95\columnwidth]{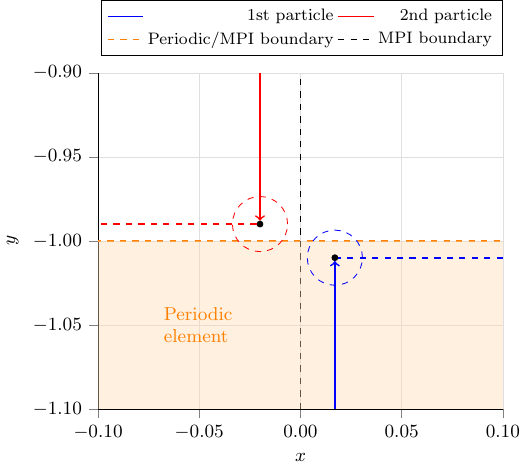}
  \caption{Sketch of the simple particle collision test case with the numerical and analytical particle trajectories including
  multiple process boundaries and periodic boundary conditions.}%
  \label{fig:validation:partcoll_simple}
\end{figure}

To assess the accuracy and robustness of the particle collision algorithm, the numerical particle collision time between two colliding particles is
compared to its analytical solution.
For the robustness check, the particles are located on different processors separated by a periodic boundary condition,
see~\cref{fig:validation:partcoll_simple}.
A computational domain of size $\Omega=[-1.1,1.1]^3$ is discretized with $4^3$ elements with periodic boundary conditions. A uniform initial solution is filled with two particles with $\partdiam=0.04$.
The first particle is located at $\partpos[p,1](t_0) = [-0.02,-0.5,0]$ with $\partvel[p,1](t_0) = [0, -15, 0]$, and the second particle is placed at $\partpos[p,2](t_0) = [0.017,0.5,0]$ with $\partvel[p,2](t_0) = [0, 15, 0]$.
The exact solution can be computed analytically by solving~\cref{eq:methods:collision:detect1,eq:methods:collision:detect2}, resulting in
$\tcoll^\text{ex}=\num{9.96237e-2}$.
The relative error between the analytically and numerically predicted time of the first particle collision is around machine precision at $\mathcal{O}(\num[scientific-notation=true,retain-unity-mantissa=false]{1e-16})$.

\subsection{Kinetic Energy Analogy and Preservation}
\begin{figure}[!tb]
  \includestandalone[width=\columnwidth]{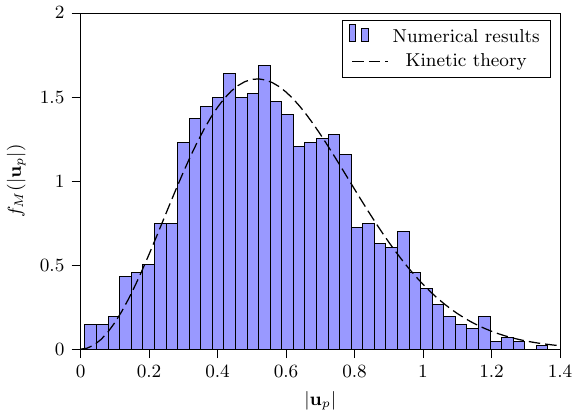}
  \caption{Distribution of the particle velocity magnitude predicted by the numerical simulation and given by the kinetic theory.}%
  \label{fig:validation:partcoll_ekin}
\end{figure}

This validation case aims to further test the robustness and efficacy of the proposed framework by checking the kinetic energy
analogy and the conservation of the total kinetic energies of the particles.
The setup is chosen such that it mimics the motion of molecules in an ideal gas under thermodynamic equilibrium, following the
principles of kinetic theory, similar to~\cite{Sundaram199775,Ching2021}. Hence, the particle distribution should converge to the three-dimensional Maxwell-Boltzmann distribution, given as
\begin{align}
  f_M\left(\abs{\partvel}\right) = \left[ \frac{2 \pi k_B T}{m} \right]^{-3/2} \exp \left( - \frac{1}{2} \frac{m \abs{\partvel}}{k_B T} \right)
\end{align}
with the Boltzmann constant $k_B$ and the equilibrium temperature $\smash{T=\partmass \overline{|\partvel|^2}(3 k_B)}^{-1}$, here $T=\num{4e9}$, where $\overline{(\cdot)}$ denotes the mean.
The computational domain is of size $\Omega=[-1,1]^3$ with periodic boundaries and is initially filled with an inviscid, quiescent fluid. The domain is equipped with \num{5000} randomly distributed particles which are initialized with $\partdiam=0.02$, $\partdens=\num[scientific-notation=true,retain-unity-mantissa=false]{1e-10}$, while the initial particle velocities are random in each direction with a constant velocity magnitude of one. This setup results in a volume fraction of around \num{4.2e-3} and about \num{2e5} particle collisions over \num{1000} time steps.
For this validation study, the particles are assumed to have no acceleration and no influence on the fluid phase, thus only particle collisions are considered.
As illustrated in~\cref{fig:validation:partcoll_ekin}, the particle velocities converge towards the three-dimensional
Maxwell-Boltzmann distribution.
Moreover, the total kinetic energy production of the particles is less than machine precision, independent of the number of
processors used, such that the total kinetic energy is preserved over time.

\section{Parallel Performance}%
\label{sec:parallel}

\elexi is designed as massively parallel code aimed towards modern high performance computing (HPC) systems.
As such, retaining scalability on large core counts is an inherent design goal.
For the current work, the scaling performance is evaluated via simulations performed on the HPE Apollo \hawk system at the High Performance Computing Center (HLRS) in Stuttgart, Germany, and on the \lumi (Large Unified Modern Infrastructure) system at the CSC–IT Center for Science in Kajaani, Finland.
\hawk utilizes dual-socket AMD EPYC\textsuperscript{TM} \num{7742} nodes (\num{128} cores per node) combined with \SI{256}{\giga\byte} RAM and an InfiniBand HDR200 interconnect in an enhanced 9D-hypercube topology.
The code was compiled with the GNU compiler version 9.2.0 with the libraries mpt 2.23, hdf5 1.10.5 and aocl 3.0.
The \lumi-C hardware partition utilizes newer AMD EPYC\textsuperscript{TM} \num{7763} nodes (\num{128} cores per node) combined with \SI{256}{\giga\byte} RAM and an HPE Cray Slingshot-11 \SI{200}{\giga\bit\per\second} interconnect with a dragonfly network topology.
Here, the code was compiled with the GNU compiler version 13.2.1 with the libraries Cray MPICH 8.1, hdf5 1.12.2 and Cray LibSci 24.03.
The scaling performance is evaluated for the time spent to advance the computational simulation, i.e., without initialization effort and input/output times.
Each run was repeated \num{5} times to eliminate fluctuations in overall machine load.
Scaling is investigated using an unstructured grid in the form of a Cartesian periodic box with elementary dimensions $2L \times L \times L, \ L \in \mathbb{R}_{>0}$ and 16 hexagonal elements/$L$.
For weak scaling runs on multiple compute nodes, this box is extended in $x_{1}$-direction through multiplication with the number of nodes to maintain identical grid spacing.
The fluid field is initialized to a uniform flow with the velocity vector $\fluidvel = [\num{1.},\num{1.},\num{1.}]^T$.
A sponge zone using an exponential temporal filter following \citet{Pruett2003} is set up in the final $x_{1}$-decile to dampen numerical oscillations before reaching the outflow boundary while periodicity constraints are imposed on the other two directions.
Particles are emitted once initially with a concentration of \num[scientific-notation=fixed,fixed-exponent=0]{0.1250e+06} particles per unit cube $L^{3}$ with their initial velocity set equal to the local fluid velocity.
First, the memory consumption and time-to-solution are discussed, followed by the parallel efficiency of \elexi.
To the author's best knowledge, this represents the first time that comprehensive performance data for one-way, two-way, and
four-way coupled Euler--Lagrange simulations is published on these state-of-the-art HPC clusters.

\subsection{Memory Consumption and Time-to-Solution}
The MPI+MPI hybrid implementation means that data required on a single compute node is stored uniquely in memory.
The resulting approach is particularly memory-efficient on Simultaneous Multi-Threading (SMT) systems with large core counts, such as common in the aforementioned HPC systems.
\Cref{tab:implementation:lumi:slurm} depicts the CPU time for \num{10} Runge-Kutta loops, summed about over all cores, and the
traceable resource total memory usage of all tasks in the job\footnote{The total memory usage is recorded by \texttt{SLURM} using the \texttt{TresUsageInTot} metric.} for a single compute node calculation on \lumi.
For the given setup, the chosen polynomials degree of $\ppn=5$ results in $\approx \num[round-mode=places,round-precision=2]{8.8473600E+05}$ degrees of freedom (DOFs) with about \SI{1.02}{\kilo\byte} per DOF, closely matching other codes of the \flexi family~\cite{Kurz2025}.
The inclusion of a 1-way coupled Lagrangian phase significantly increases memory consumption.
Note that only a minor portion of this memory is allocated is devoted to storing the actual particle data with about \num{350} bytes per particle~\cite{Kopper2023}.
Most of the additional memory is dedicated towards providing the geometric information necessary for particle tracking.
Enabling of particle-to-fluid interaction increases the memory further, as information for each unique deposition node must be available.
From this point on, our proposed MPI+MPI hybrid approach for particle collision only adds minor memory consumption.
However, the exact calculation of the particle interactions results in a runtime increase with a factor of $\approx \num{8.3}$.
\begin{table}[!tb]
  \begin{tabularx}{\linewidth}{lc*{1}{>{\centering\arraybackslash}X}}
  \toprule
  Coupling           & CPU Time [\si{\second}]                          & Memory [\si{\mega\byte}]                         \\\midrule
  0-way (pure fluid) & \makebox[\widthof{\num{150.55}}][r]{\num{5.50}}  & \makebox[\widthof{\num{5819.6}}][r]{\num{902.9}} \\
  1-way              & \makebox[\widthof{\num{150.55}}][r]{\num{11.52}} & \num{5819.6}                                     \\
  2-way              & \makebox[\widthof{\num{150.55}}][r]{\num{18.12}} & \num{6438.7}                                     \\ %
  4-way              & \num{150.55}                                     & \num{6517.8}                                     \\\midrule
  \bottomrule
  \end{tabularx}
  \caption{CPU time and memory consumption for $\num[round-mode=places,round-precision=2]{8.8473600E+05}$ degrees of freedom and $\num[round-mode=places,round-precision=1]{2.4999400E+05}$ particles for different degrees of phase coupling on one compute node on \lumi.}%
  \label{tab:implementation:lumi:slurm}
\end{table}

\subsection{Parallel Efficiency}
\begin{figure}[!tb]
  \centering
  \includestandalone[width=\columnwidth]{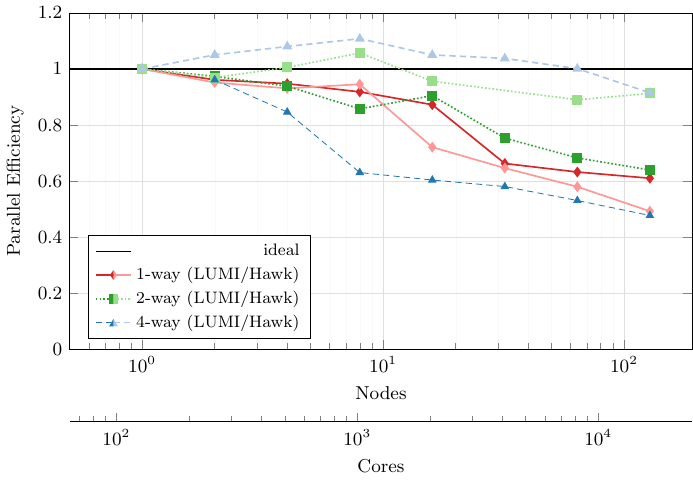}
  \caption{Weak scaling of \elexi with the split-form DG scheme and $N=\num{5}$ plotted as the parallel efficiency over the number of CPUs for fixed loads, i.e., DOF/Particles per CPU. The parallel efficiency is computed based on the performance on a compute single node, i.e., on 128 CPUs.}%
  \label{fig:scaling:weak}
\end{figure}

Weak scaling efficiency of \elexi for one-way, two-way, and four-way coupling between the phases is presented in \cref{fig:scaling:weak}.
As the MPI-3 shared memory parallelization is performed on a compute node level, the performance is normalized relative to the computational capability of a single compute node, consisting of \num{128} CPUs.
The only exception is the four-way coupled simulation on \lumi, where the additional MPI windows exhaust the hardware message queue of the Slingshot-11 interconnect.
This exhaustion occurs for all four-way coupled simulations using two or more \lumi nodes, requiring a transition to slower software message matching.
These cases were normalized to align with the one- and two-way coupled results for two nodes on \lumi to ensure consistency.

One-way coupled simulations have the lowest compute-to-communication ratio, rendering them the most sensitive to bandwidth and latency limitations imposed by the interconnect.
Weak scaling efficiency consistently exceeds \SI{61}{\percent}, demonstrating reliable performance even at the scale of \num{128} nodes (\num{16384} cores) on \lumi.
As the Slingshot-11 interconnect in dragonfly topology on \lumi features a lower diameter compared to the InfiniBand network with a hypercube topology on \hawk, one-way coupled performance results on \hawk fall below similar simulations on \lumi in terms of efficiency for high node counts.
In contrast, weak scaling results for two- and four-way coupling on \hawk exhibit a performance increase up to \num{8} nodes, which was confirmed to originate from the energy-optimized runtime environment provided by the PowerSched framework~\cite{Simmendinger2024}.
At higher node counts, simulation performance on \hawk is again limited by the interconnect which results in a drop of weak scaling efficiency.
However, the increased computational load compared to the one-way coupled approach result in \SI{91.5}{\percent} weak scaling efficiency for the four-way coupled simulation on \num{128} \hawk nodes.
The greater computational capacity of the \lumi nodes, combined with the challenges associated with MPI message matching, results in reduced weak scaling efficiency at high node counts for the same test case on \lumi.
Higher node counts exceed the $2^{31}$-limit of \num{4}-byte signed integers and are not considered for the weak scaling performance.

\begin{figure}[!tb]
  \centering
  \includestandalone[width=\columnwidth]{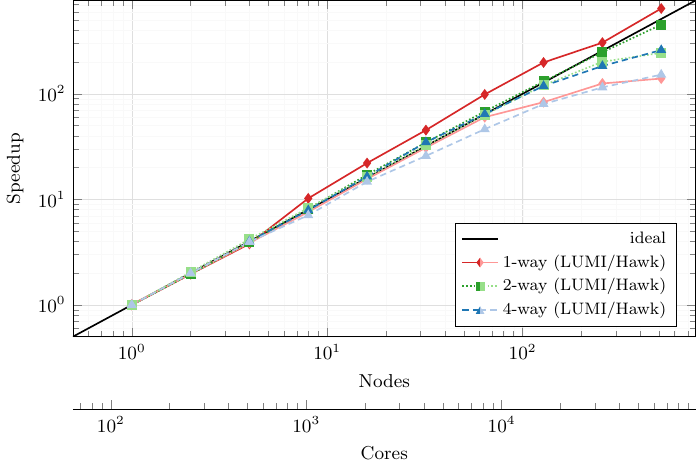}
  \caption{Strong scaling of \elexi with the split-form DG scheme and $N=\num{5}$ plotted as the parallel efficiency over the number of CPUs for fixed case size, i.e., total number of DOF/Particles. The speed-up is computed based on the performance on a compute single node, i.e., on 128 CPUs.}%
  \label{fig:scaling:strong}
\end{figure}

Strong scaling results in form of the computational speed-up for the $128L \times L \times L$ case, thus approx \num[exponent-mode=scientific,round-mode=figures,round-precision=3]{5.6623104e+07} DOFs and \num[exponent-mode=scientific,round-mode=figures,round-precision=3]{1.6000116E+07} particles, are depicted in \cref{fig:scaling:strong}.
Similar to the weak scaling tests, all tests were performed for one-way, two-way, and four-way coupling between the phases.
Exhaustion of the Slingshot-11 hardware message queue occurs at \num{4} \lumi nodes which was thus chosen as reference for normalization.
For the one-way and two-way coupled simulations on \hawk, nearly ideal speed-up is observed up to \num{64} compute nodes (\num{8192} Cores).
At higher core counts, computational jobs at \hawk are transferred from a dedicated partition to distributed racks using best-effort topology-aware scheduling.
This leads to a gradual decrease in strong scaling efficiency as jobs with $\geq 128$ compute nodes operate with higher interconnect latency due to the hypercube topology.
The effect is more pronounced for the one-way coupled case as the deposition of the particulate force on the fluid domain is performed with an expensive atomic operation in the time-stepping loop.
While this atomic operation could be circumvented by allocating shadow arrays, we prioritized optimizing \elexi towards lower memory consumption in this space-time/time-memory trade-off.
Strong scaling efficiency for four-way coupled simulations initially remains below those neglecting collisions as there is limited computational load available to hide the MPI RMA operations.
While this reduces the speed-up at low core counts, the one-sided nature of these operations translates into improved scalability compared to one/two-way coupled simulations at high node numbers, indicating potential for further scaling.
Results on \lumi generally outperform the corresponding simulations on \hawk, with the one-way coupled simulation even demonstrating superscalar scaling due to reduced memory pressure.
Both the one-way and the two-way coupled simulation on \lumi show slopes parallel to ideal scaling up to the maximum of \num{512}
compute nodes (\num{65536} cores) tested while the four-way coupling approach enters the area of diminishing returns for $>
\num{256}$ compute nodes.
Overall, the results on both \lumi and \hawk affirm the scalability and adaptability of \elexi across diverse hardware environments, enabling efficient calculation of increasingly complex simulations at scale.

\section{Applications}%
\label{sec:application}
In this section, the applicability of \elexi to more challenging large-scale test cases is demonstrated.
For this, a plane particle-laden jet impinging on a cavity and the particle-laden flow around a transonic NACA0012 airfoil under atmospheric
conditions comparable to the Martian environment are considered.
In the following, only viscous forces, i.e., drag and heat, are taken into account.
In all cases, four-way coupling between the fluid and the dispersed phase is assumed.
Finally, we want to emphasize that the current applications only serve as numerical examples.
Thus, detailed physical investigations are out of the scope of this paper.

\subsection{Round Jet Impinging on a Cavity}
In this section, we investigate the particle behavior for a plane particle-laden jet impinging on a cavity by comparing a
simulation without particles to a simulation with a four-way coupled dispersed phase.
The primary focus of this application is to demonstrate the capabilities of \elexi as well as the effectiveness of the load
balancing.
The setup is chosen to mimic a dry-ice blasting procedure, where a high-velocity jet laden with frozen carbon dioxide particles is utilized to
clean solid surfaces. Due to the various physical phenomena that occur during the process, this test case can be considered particularly challenging~\cite{Spur1999}.
For the chosen setup, the continuous phase is air with a dynamic viscosity of $\mu=\SI{2.71e-5}{\kilogram\per\meter\per\second}$. The jet enters the domain with a Mach number of $M = \num{0.6}$ and a Reynolds number of $Re=89605$
based on the jet radius $r_\text{jet}=\SI{0.05}{\meter}$ and the characteristic flow velocity of the jet $u_\text{jet}=\SI{275.0679}{\meter\per\second}$.
The computational domain is discretized using \num{392630} hexahedral elements with $\ppn=4$.
The domain is defined as $\smash{\Omega=\Omega_\text{box} \cup \Omega_\text{cavity}}$ with $\smash{\Omega_\text{box} \in [0,0.0333] \times [0,0.06] \times [0,0.0008] \mskip3mu \si{m}}$
and a cavity $\smash{\Omega_\text{cavity}=[0.0333,0.03665] \times [0,0.015] \times [0,0.0008] \mskip3mu \si{m}}$.
Adiabatic no-slip boundary conditions are prescribed on the upper domain, while pressure outflow conditions
($p=\SI{1.2e5}{\pascal}$) are set at the left and
lower domain. Inflow conditions are defined based on the total pressure and total temperature of the jet, $p_t=\SI{144097}{\pascal}$ and
$T_t=\SI{331.8}{K}$, respectively, at the lower right domain. Periodic boundary conditions are prescribed in the third direction.
The fluid is initially at rest with $p=\SI{1.2e5}{\pascal}$ and $\rho=\SI{1.2597}{\kg\per\cubic\meter}$.
Following \citet{Liu2012}, particles are emitted with a size of $\smash{\partdiam = \{20,28,40,57,80\} \mskip3mu \mu \si{\meter}}$, a density of $\smash{\partdens = \SI{1560}{\kilogram\per\cubic\meter}}$, an initial temperature
of $\smash{T_p = \SI{195}{\kelvin}}$, and a specific heat at constant pressure of $c_p =
\SI{519.16}{\joule\per\kelvin\per\kilogram}$. The emission rate of the particles is chosen such that a volume fraction of
$\approx \num{1e-3}$ is reached in the jet.
The simulation is carried out without particles until $t=4T^\ast$, with the characteristic time
$T^\ast=L_\text{jet}/u_\text{jet}$ defined as the ratio of the length of the jet $L_\text{jet}=\SI{0.03665}{\meter}$ to the characteristic velocity of the jet.
Without load balancing, the simulation achieved a time-to-solution of \SI{1.1e-3}{s} per degree of freedom and Runge–Kutta stage, which was nearly halved to \SI{5.6e-3}{s} through dynamic workload rebalancing.

\begin{figure*}[htb]
  \centering%
  \includestandalone[width=\linewidth]{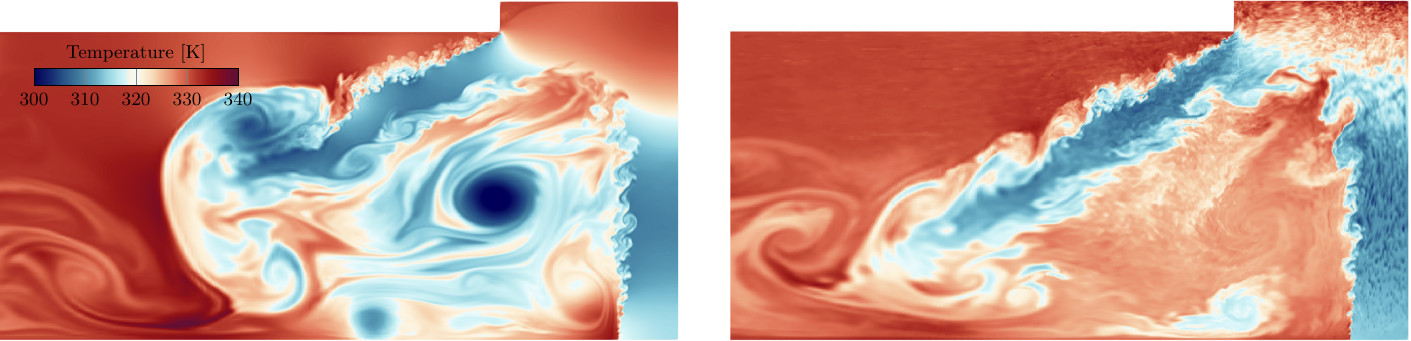}
  \caption{Instantaneous temperature distribution of a round jet impinging on a cavity at $4T^\ast$ without (left) and at $8T^\ast$ with 4-way coupled
  particles (right).}%
  \label{fig:application:partjet}
\end{figure*}

Slices of the corresponding temperature profile without and with four-way coupled particles at simulation times of $t=4,8 \mskip3mu T^\ast$ are depicted in~\cref{fig:application:partjet}.
In the simulation without particles, a second jet emerges due to the chosen size and depth of the cavity, which, combined with the main jet, forms a strong recirculation area comprised of a primary vortex and various small vortices.
As illustrated on the right of~\cref{fig:application:partjet}, the presence of a four-way coupled particulate phase strongly
modifies the flow field. First, the influence of the dry-ice particles is clearly visible in the locally decreasing temperature field. Second, the mean characteristics of the complex flow are slightly stabilized due to the viscous particle forces imposed on the continuous phase.
Simultaneously, the high number of particle collisions causes local perturbations.
Thus, neglecting particle collisions in dense particle-laden flows can lead to vastly different flow structures
and in turn particle motions.

\subsection{Airfoil Flow under Martian Atmospheric Conditions}
This test case was chosen to demonstrate the performance of \elexi at scales representative of practical applications.
The flow around a NACA0012 airfoil under atmospheric conditions similar to the Martian environment was investigated experimentally~\cite{Nguyen2020} and numerically~\cite{Liu2023}.
\elexi was validated against the experimental results with no suspended particles at $M = \num{0.6}$ and $Re = \num{50000}$ with the results summarized in \cref{tab:application:naca0012:validation}.
\begin{table}
  \begin{tabularx}{\linewidth}{lc*{2}{>{\centering\arraybackslash}X}}
  \toprule
  Parameter        & Variable                                  & Experiment    & \elexi\\\midrule
  Spec.\ gas cons. & $R$ [\si{\joule\per\kilogram\per\kelvin}] & -             & \num{191.14}         \\
  Heat cap.\ ratio & $\gamma$ [-]                              & -             & \num{1.351}          \\
  Temperature      & $T$ [\si{\kelvin}]                        & -             & \num{213}            \\\midrule
  Lift coeff.      & $C_l$ [-]                                 & \num[round-mode=places,round-precision=4]{0.4204886255838629} & \num[round-mode=places,round-precision=4]{0.397191} \\
  \bottomrule
  \end{tabularx}
  \caption{Flow field parameters and comparison of the LES results of \elexi against experiments~\cite{Nguyen2020} for the flow without suspended particles.}%
  \label{tab:application:naca0012:validation}
\end{table}
Atmospheric parameters were selected according to \citet{Rafkin2020} and the angle of attack was kept stationary at $\alpha = \SI{5}{\degree}$.
For Mach numbers exceeding \num{0.7}, URANS simulations with the transition SST model and a coupled discrete phase model performed by \citet{Liu2023} found that sand particles contained in the Martian atmosphere break up the stable shock structures observed in the unloaded flow.
Consequently, the current study employs the LES approach with an inflow Mach number $Ma=\num{0.8}$ and Reynolds number $Re=\num{50 000}$.
Adiabatic no-slip boundary conditions are set on the airfoil surface with the far field prescribed by Dirichlet boundary conditions and a periodicity constraint applied in the spanwise direction.
The domain is discretized with \num{68000} hexahedral elements and a polynomial degree $\ppn=4$, resulting in \num{8.5E+06} degrees of freedom.
The simulation was performed on \num{16} \hawk nodes with a simulation efficiency (simulated physical time per utilized core hour) of \SI[round-mode=places,round-precision=2]{5.76028E-08}{\second\per CPU\hour} and allowed to run until the integral forces on the airfoil were statistically stationary.

\begin{figure*}[htb]
  \centering%
  \begin{subfigure}{0.47\linewidth}
    \includestandalone[width=\linewidth]{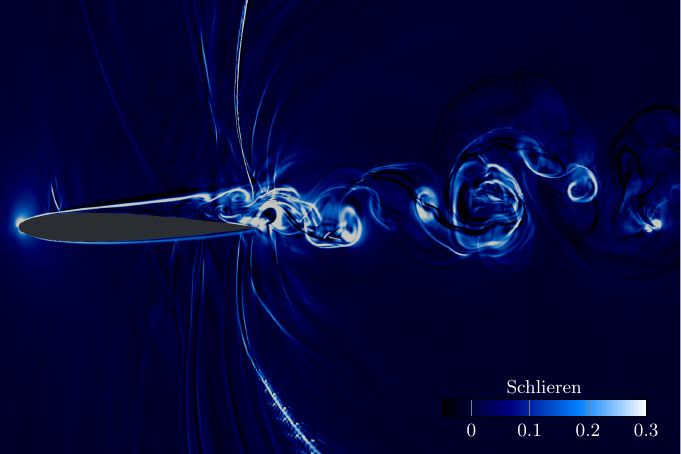}
  \end{subfigure}
  \hfill%
  \begin{subfigure}{0.47\linewidth}
    \includegraphics[width=\linewidth]{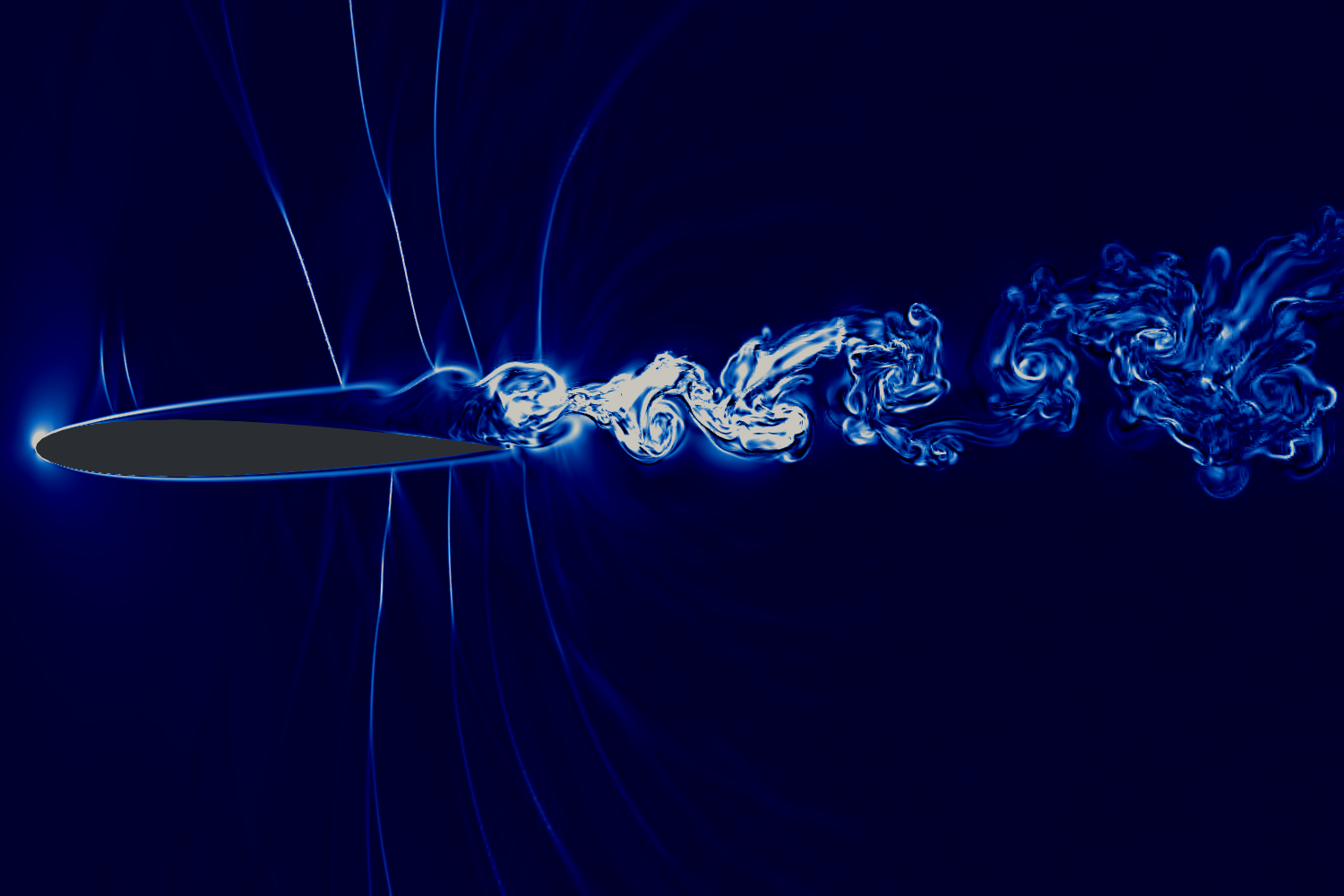}
  \end{subfigure}
  \caption{Comparison of pseudo-Schlieren as a representation of density gradients for the unladen (left) and particle-laden (right) flow around a NACA 0012 airfoil at $Ma=\num{0.8}$ and $Re=\num{50 000}$.}%
  \label{fig:application:naca0012:particle}
\end{figure*}

A comparison of a pseudo-Schlieren visualization of the resulting instantaneous LES flow field for the simulation without particles and the case with a sand particle concentration of \SI{225}{\milli\gram\per\meter\cubed} is shown in \cref{fig:application:naca0012:particle}.
The instantaneous snapshots of both simulations exhibit a shock on the forward suction side, followed by flow separation and a second shock system near the recirculation region at the trailing edge.
While the main shock front on the pressure side is located near the trailing edge for the unladen flow, this shock is pushed forward in the presence of particles.
A similar shock is also observed in the particle-laden RANS simulations conducted by \citet{Liu2023}, but their time-averaged results do not predict a shock formation on the pressure side for the unladen case.
Moving downstream, the recirculation flow develops into an irregular vortex street.
Here, the augmentation of turbulent fluctuations, introduced by the inertial particles, is clearly evident.
At the same time, the downstream momentum introduced by the particulate phase re-energizes the wake, resulting in a reduction in wake width.
These results highlight the importance of time-accurate simulations, as the presence of inertial particles significantly alters the instantaneous flow structures and reveals that stationary flow solution neglect significant parts of the reult.

\section{Conclusion}%
\label{sec:conclusion}
The efficient and accurate numerical treatment of dense particle-laden flows is challenging, especially on high performance
computing (HPC) systems.
Focusing on literature for Euler--Lagrange particle tracking in a compressible carrier fluid, the primary focus is placed more on the time-accurate
coupling of both phases and the adequate collision treatment than on the efficiency on highly parallel systems.
This work aimed to alleviate this deficiency by proposing a four-way coupled Euler--Lagrange approach based on the combination of
the particle operator with an MPI+MPI hybrid approach.
The proposed algorithm enables a highly efficient and accurate calculation of binary inter-particle collisions including the
effective treatment of the particle-fluid coupling in a compressible carrier phase on arbitrary core counts.
Special focus was placed on the detection of the particle collisions on parallel systems with possibly curved element faces.
Built on pure MPI following the MPI-everywhere strategy, the approach offers flexibility and portability across various systems, requiring no modifications for varying CPU counts per compute node.
The implementation is thoroughly validated and the excellent scaling properties on massively parallel systems are demonstrated.
This work concluded with two more challenging test cases to demonstrate its applicability to large-scale applications.
Finally, the work is open-source available and welcomes external contributions.

\section*{Acknowledgements}
The research presented in this paper was funded in parts by Deutsche Forschungsgemeinschaft (DFG, German Research
Foundation) under Germany's Excellence Strategy - EXC 2075 - 390740016 and by the European Union.
This work has received funding from the European High Performance Computing Joint Undertaking (JU) and Sweden, Germany, Spain, Greece, and Denmark under grant
agreement No 101093393.
We acknowledge the support by the Stuttgart Center for Simulation Science (SimTech).
The authors gratefully acknowledge the support and the computing time on \hawk provided by the HLRS through the project
``hpcdg''.
We acknowledge the EuroHPC Joint Undertaking for awarding this project access to the EuroHPC supercomputer \lumi, hosted by CSC
(Finland) and the LUMI consortium through a EuroHPC Development Access call.

\bibliographystyle{elsarticle-num-names}
\bibliography{references}

\end{document}